# Fabrication of Artificial Graphene in a GaAs Quantum Heterostructure

Running authors: Diego Scarabelli, Sheng Wang, Yuliya Y. Kuznetsova, Loren N. Pfeiffer, Ken West, Geoff C. Gardner, Michael J. Manfra, Vittorio Pellegrini, Aron Pinczuk and Shalom J. Wind

Diego Scarabelli, Sheng Wang, Aron Pinczuk, Shalom J. Wind

Columbia University, Department of Applied Physics and Applied Mathematics, 500 W. 120th St., Mudd 200, MC 4701, New York, NY 10027

Yuliya Y. Kuznetsova

Columbia University, Department of Physics, 538 W. 120th St., 704 Pupin Hall MC 5255, New York, NY 10027

Loren N. Pfeiffer, Ken West

Princeton University, Department of Electrical Engineering, Olden Street, Princeton, NJ 08544

Geoff C. Gardner, Michael J. Manfra

Purdue University, Department of Physics and Astronomy, and School of Materials Engineering, and School of Electrical and Computer Engineering, 525 Northwestern Avenue
West Lafayette, IN 47907

Vittorio Pellegrini

Istituto Italiano di Tecnologia, Graphene Labs, Via Morego 30, I-16163 Genova, Italy
NEST, Istituto Nanoscienze-CNR and Scuola Normale Superiore, Piazza San Silvestro 12, I-56127 Pisa, Italy

Electronic mail: sw2128@columbia.edu

The unusual electronic properties of graphene, which are a direct consequence of its two-dimensional (2D) honeycomb lattice, have attracted a great deal of attention in recent years. Creation of artificial lattices that recreate graphene's honeycomb topology, known as artificial graphene, can facilitate the investigation of graphene-like phenomena, such as the existence of massless Dirac fermions, in a tunable system. In this work, we present the fabrication of artificial graphene in an ultra-high quality GaAs/AlGaAs quantum well, with lattice period as small as 50 nm, the smallest reported so far for this type of system. Electron-beam lithography is used to define an etch mask with honeycomb geometry on the surface of the sample, and different methodologies are compared and discussed. An optimized anisotropic reactive ion etching process is developed to transfer the pattern into the AlGaAs layer and create the artificial graphene. The achievement of such high-resolution artificial graphene should allow the observation for the first time of massless Dirac fermions in an engineered semiconductor.

## I. INTRODUCTION

Charge carriers in graphene are said to behave as photon-like relativistic particles described by the Dirac equation, with null rest mass and constant Fermi velocity $v_F = 10^6 \ m/s$, known as massless Dirac fermions (MDFs)[1, 2, 3, 4]. This peculiar behavior, deriving from the triangular symmetry of the honeycomb lattice, has spawned a novel field of condensed matter science and technology aimed at investigating and controlling the relativistic character of Dirac quasi-particles. Artificial honeycomb lattices, also known as "artificial graphene" (AG), have been proposed as a potential platform for creating and manipulating Dirac bands in a highly tunable manner[5, 6], allowing access to quantum regimes that cannot be attained in natural graphene, whose lattice structure is fixed and is difficult to modify. By adjusting the design of the artificial lattice it should be possible to control the lattice parameter and the inter-site coupling, with effects on the Fermi velocity of the MDFs and on the inversion symmetry of the lattice, the removal of which would introduce a band gap. It would also be possible to intentionally introduce defects and edges to modulate the band structure. In addition, a larger lattice parameter accommodates a higher magnetic flux per unit cell, creating a regime where the magnetic length and the lattice period are comparable. In these difficult to reach conditions the interplay between magnetic and electronic states are predicted to create a fractal energy spectrum that could lead to the observation of exotic physics such as Hofstadter's butterfly[7].

Thus far, AG physics has been observed in molecular[8] and optical lattices[9]. Particularly promising would be the realization of AG in an engineered high-mobility semiconductor system, as conventional fabrication methods could provide the means for a more adjustable, scalable system, as well as one that is easier to integrate with other electronic components. A two-dimensional electron gas (2DEG) whose electron density is modulated by a periodic electrostatic potential with honeycomb topology has been predicted to host tunable MDFs[10, 11]. The creation of such a honeycomb potential by lithographically defining a lateral superlattice on a GaAs heterostructure hosting a 2DEG has recently been demonstrated[12, 13, 14]; however, direct evidence of MDFs has not yet been obtained in this type of system, primarily because the relatively large lattice parameter in those devices (> 120 nm) cannot create energy minibands sufficiently large to overcome the effects of temperature and disorder. In order to observe AG effects in a 2DEG system, it is essential to increase the energy range where the linear dispersion occurs by reducing the AG lattice period and advancing its spatial uniformity. In this work we report the realization of AG lattices with period as small as 50 nm in an ultra-high mobility 2DEG, obtained by performing a shallow dry etch of the sample surface masked by a honeycomb array of metallic nano-disks defined by electron-beam lithography and metal deposition.

## II. EXPERIMENTAL

The heterostructure used in this work, shown schematically in Fig. 1(a), was designed to create a 2DEG confined in a 25 nm-wide one-side modulation-doped GaAs/Al$_{0.1}$Ga$_{0.9}$As quantum well (QW), grown by molecular beam epitaxy. The Si δ-doping layer is 80 nm below the surface of the sample and 30 nm above the 2DEG. The measured as-grown electron density was 1.8 x 10$^{11}$ cm$^{-2}$ and the low-temperature mobility was 3.2 x 10$^6$ / (Vs).

A periodic potential modulation of the 2DEG can be obtained by either patterning a metallic layer deposited onto the sample used as a top gate[15, 16, 17, 18] or by etching the sample surface, creating an array of nano-dots [11, 12, 13, 14, 19, 20]. The first approach presents challenges in the control of the potential amplitude perceived by the electrons of the 2DEG, which decreases with increasing distance from the gate. The second method consists of effectively bringing the surface closer to the 2DEG so that more electrons tend to occupy the surface states; this causes a local decrease in the electron density beneath the etched areas, as illustrated in Fig. 1(b) for the case of a 2DEG in a modulation-doped GaAs/AlGaAs quantum well, the system used in this work. The pillars resulting from the masked dry etch act as an effective attractive potential for the electrons, forming a honeycomb lattice of quantum dots in the 2DEG. This approach is both simpler to implement and more effective than gating the sample surface.

A detailed account of the fabrication of the artificial honeycomb lattices is reported in the next sections. In part A we describe different approaches for the creation of a mask for the dry etching of the quantum well surface. Electron-beam lithography at 80 kV (NanoBeam *n*B4) was used to pattern honeycomb arrays of dots in one case (1) on hydrogen silesquioxane (HSQ), a negative tone resist, a simpler process for the creation of the mask. An alternative process (2) uses polymethyl-methacrylate (PMMA), a positive tone resist, followed by metal deposition and lift-off, to form a metallic mask. A variation on this last method is also presented in (3). In part B we describe the process developed for the inductively-coupled plasma reactive-ion etching (ICP-RIE), based on a gas mixture containing $BCl_3$ and Argon.

## A.  Fabrication of the etching-mask with honeycomb topology

### 1.  Direct patterning of HSQ

Samples were spin-coated with a diluted HSQ solution (XR1541-006 : methyl isobutyl ketone = 1 : 1) at 4000 rpm for 45 s to produce a 50 nm-thick film. One of the samples was pre-treated with SurPass3000 (DisChem), a waterborne cationic adhesion promoter, prior to the resist coating. Electron beam exposure of the resist was carried out with a beam current of 300 pA. Honeycomb arrays of circles of diameter 75 nm and 25 nm with pitch 150 nm and 50 nm, respectively, were exposed at electron doses ranging from 1000 $\mu C/cm^2$ to 2000 $\mu C/cm^2$ in an exposure area of 50 μm x 50 μm. The resist was developed in Microposit MF CD-26 (TMAH 2.6% aqueous solution) at room temperature for 4 min and rinsed for 2 min in boiling ethanol.

## 2. Patterning PMMA, metal deposition and lift-off

For this process, a double layer resist stack of polymethyl-methacrylate (PMMA) was employed. The process flow is illustrated in Fig. 2(a-d). The sample was spin-coated with a 40 nm-thick film of PMMA of molecular weight 35 K, to form the bottom layer, and then baked for 5 min at 180 ºC. Another 30 nm thick layer of 495 K molecular weight PMMA was spun and baked at 180 ºC for 15 min. The sensitivity of PMMA is higher for lower molecular weight, consequently the bottom layer develops more rapidly than the top layer, causing an under-cut profile to form upon development that facilitates the subsequent metal lift-off. The e-beam exposure was performed with beam current 400 pA and accelerating voltage 80 kV. The 200 μm x 200 μm AG lattices were patterned by exposing honeycomb arrays of circles varying the lattice period (40, 50, 60, and 70 nm); the diameter of the circles was varied as a function of period, and the exposure dose ranged between 500 μC/cm$^2$ and 2500 μC/cm$^2$. The resist was developed for 60 s in a solution of methyl-isobutyl ketone : isopropanol (MIBK : IPA, 1 : 3 by volume) at 5 ºC, applying ultrasonic agitation for increased resolution and contrast[21, 22], then rinsed in isopropyl alcohol. After the development the samples underwent a 10 s oxygen plasma treatment (Diener Tetra 30 PC plasma cleaner, 0.25 mbar, 4 sccm O$_2$ flow, 212 V DC bias, 300 W RF power at 13.56 MHz) meant to eliminate residual resist from within the developed features and terminate the surface of the sample with hydroxyl groups, which can aid in adhesion of the metal mask.

Electron beam evaporation was used to deposit 2 nm Ti (adhesion layer) followed by 8 nm Au. The metal deposition was performed at a rate of 0.5 Å/s while maintaining the chamber pressure below 5 x 10$^{-7}$ mbar (Angstrom EvoVac e-beam evaporator). The lift-off of the metal was accomplished by soaking the sample for 12 hours in Remover PG (MicroChem), solution containing N-Methyl-2-Pyrrolidone (NMP), and then spraying acetone with a syringe for complete removal of the resist and any metallic debris, followed by rinsing in IPA and water.

### 3. Hard-mask and oxygen plasma descum

A variation on the previous method was developed in order to improve the lift-off conditions and to obtain disks with smaller diameter, as discussed in more detail in the "Results and Discussion" section. In this process, samples were coated with a single layer of PMMA 495 K, 60 nm thick, that was exposed and developed as described above. After development, a 5 nm thick layer of Ti was deposited at an angle of 60º with respect to the vertical (angled evaporation, Angstrom EvoVac) in order to form a metallic hard-mask on the surface of the resist without reaching the bottom of the developed features (Fig. 3). The hard-mask protects the resist during the $O_2$ RIE descum (Oxford PlasmaLab 80 Plus ICP 65), which also renders the surface hydrophilic and increases the adhesion of the deposited metal. The hard mask also helps form an under-cut profile which facilitates the lift-off. The descum was performed at 80 sccm $O_2$ flow, chamber pressure 0.05 mbar, 60 W RF power at 13.56 MHz, DC bias 210 V, for 30 s. Deposition and lift-off were done as described above.

## B. Pattern transfer by reactive ion etching

The honeycomb array of HSQ nano-dots or Au nano-disks was transferred to the substrate using inductively-coupled plasma reactive-ion shallow etching (ICP-RIE) (Samco International RIE200iP, PRISM Micro/Nano Fabrication Laboratory at Princeton University). The plasma conditions were controlled by means of two independent RF power sources. The first RF source (RF1) was inductively coupled to the gas in the process chamber (13.56 MHz, maximum power 1000 W), capable of generating a high density of radicals and ions without applying a potential difference between the plasma and the wafer surface. The second RF source (RF2) induces a DC bias between the chuck and the plasma (13.56 MHz, maximum power 300 W), as in the case of traditional RIE, primarily affecting the ion energy. The advantages of using ICP etching over conventional parallel plate RIE are well documented; of particular importance to this work is the ability to adjust the plasma density and the ion energy separately, allowing for a better tuning of the degree of anisotropy of the etch. Samples were mounted on a silica carrier wafer with silicone-based vacuum grease, and the chuck temperature was maintained at 50 ºC during the etch. The gas employed was a mixture of 50 sccm Ar and 5 sccm $BCl_3$, which has been shown to achieve non-selective etching of GaAs/AlGaAs heterostructures with good degree of anisotropy[23, 24]. We independently varied the power of the two RF sources within the range 50 - 75 W in order to optimize the vertical sidewall profile. The chamber pressure was set at 3.7 mbar in all the cases, while the etch time was varied between 60 s and 110 s.

## III. RESULTS AND DISCUSSION

The first approach, based on the use of HSQ, was limited by the lack of resist adhesion to the substrate. As displayed in Fig. 4(a-d), reasonably good adhesion could be achieved in the case of a 150 nm period honeycomb array of HSQ nano-dots with 75 nm diameter when a cationic adhesion promoter was used. However, when the lattice period and the dot diameter were reduced to 50 nm and 25 nm respectively, resist adhesion failure increased, and only overdosed dots remained on the sample surface without being moved or washed away during the resist development, as shown in Fig. 4(c, d). The inability to maintain robust adhesion with the smaller diameter dots prevented this method from reaching the lattice scale required to create artificial graphene.

The second method, which creates a metallic etch mask via lift-off, was shown to be capable of achieving honeycomb lattices with period as small as 40 nm and with excellent uniformity. Figure 5(a-d) shows representative SEM micrographs of Au masks with lattice period 50 nm and 70 nm and different disk diameters. The ability to vary the disk diameter is critical for optimal tuning of the honeycomb potential that generates the artificial graphene. Moreover, in reducing the scale it is essential to maintain a high lattice uniformity, i.e., homogeneous disk diameter and lattice period, as well to avoid lattice defects, such as missing or misplaced sites. Lower magnification SEM images of honeycomb lattices with 40 nm and 50 nm period can be seen in Fig. 5(e) and (f) respectively, for which these requirements are met over dozens of unit cells.

We note that the achievement of high-resolution and uniform arrays over large areas (up to 200 x 200 $\mu m^2$ in this work) by electron beam lithography and lift-off becomes increasingly challenging as the feature size and period are decreased. This is especially true for high atomic number substrates such as GaAs (Z=32), where the backscattering intensity is ~ 40% higher that on Si[25]. The process we have developed is both high resolution and low contrast, enabling us to achieve uniform honeycomb lattices with periods smaller than 50 nm.

After etching, the samples were examined in a Hitachi S-4700 SEM, with 65º tilt angle, acceleration voltage 20 kV and current 10 µA. The two AG lattices shown in Fig. 6 had identically patterned hard-masks but were etched with two different processes, as described above in Section II. One can see how the lattice in Fig. 6(a), which underwent an etch with higher ICP RF power (75 W) and lower bias-inducing RF power (50 W), presents evidence of lateral etching, resulting in the collapse of some pillars. Modification of the process lead to an optimal combination of ICP and bias-inducing RF powers (50 W and 75 W respectively) which resulted in straight sidewalls as shown in Fig. 6(b). The increase of the physical bombardment due to the higher ion energy, and the reduced chemical etching due to a lower plasma density, did not significantly alter the selectivity of the RIE process; in fact, in both cases the gold nano-disks are still observable as brighter dots on top of the etched pillars. It is worth noting that in all cases we observed a reduction in the diameter of the metallic nano-disks overlying each pillar following the etch. SEM images are presented in Fig. 7, which shows 50 nm period AG lattices patterned for different etching times, from 60 s to 110 s, using the optimized etch process. The height of the pillars determines the intensity of the effective periodic potential and the final degree of modulation of the electron density of the 2DEG; therefore it is important to control the etch depth with nanometer accuracy. For 110s etch time, shown in Fig. 7(d), the measured depth was ~75 nm, only 5 nm above the Si δ-doping layer. This should give the strongest potential modulation and well-defined AG minibands with linear dispersion. The verticality of the etch profile decreases with increasing etch time; nevertheless the mechanical stability of the pillars appears to be unaffected. The aspect ratio of the fabricated pillars in Fig. 7(d) is 3.5.

As noted above, the challenges related to proximity effects and the reliability of the lift-off can be managed using the hard mask method described in the experimental section. The hard mask actually allows a slight underexposure of the honeycomb pattern, resulting in minimal broadening of the features, increasing the process window and the ultimate resolution achievable. The $O_2$ plasma etching process produces an undercut in the resist that eases the lift-off. Furthermore, the complete removal of the resist residue and the hydrophilization of the surface facilitates the adhesion of the metal, promoting defect-free arrays with excellent long range order. A disadvantage of the hard mask method is an increase in the edge roughness of the metal nano-disks, resulting from the irregularity of the grains of the metal coating the border of the developed resist features. This can affect the uniformity of the lattice, as can be seen in Fig. 8(a) for a 50 nm period array. A possible way to improve on this aspect is to anneal the sample, causing each metal disk to melt and form a smoother droplet, as displayed in Fig. 8(b) and (c), in which case the annealing has been performed on a hot plate at 450 °C for two hours. Another consequence of the use of a metal hard mask and high temperature annealing is that the nano-disks are smaller. The hard mask in fact narrows the diameter of the apertures, and the annealing turns disks into nanoparticles, shrinking their in-plane cross section. This method can be used to achieve very narrow pillars, as can be seen in Fig. 8(d), where the average width is about 10 nm, with aspect ratio 5.

## IV. SUMMARY AND CONCLUSIONS

High resolution electron beam lithography and reactive ion etching were used to create artificial honeycomb lattices with period as small as 50 nm on GaAs/AlGaAs quantum wells hosting a two dimensional electron gas. The aim is to replicate in a tunable manner the massless Dirac fermion physics which is a hallmark of graphene. The direct fabrication of the mask by employing HSQ suffered from issues related to the adhesion of the resist to the substrate that are difficult to overcome. The use of bilayer PMMA, metal deposition and lift-off enables the achievement of patterns with high resolution, excellent uniformity and long range order. By performing a post-development oxygen plasma descum protected by a metal hard mask, and post-processing annealing of the etch mask, it is possible to obtain AG lattices with higher aspect ratio. The optimized etch process yields nearly vertical sidewalls, with no evidence of pillar collapse. By probing the electron states of these devices we expect to be able to observe for the first time a graphene-like band structure in an artificial semiconductor lattice.

## ACKNOWLEDGMENTS

Major activities in this project were supported by grant DE-SC0010695 funded by the US Department of Energy Office of Science, Division of Materials Sciences and Engineering. Fabrication and processing at Columbia were performed in the CEPSR Cleanroom under the auspices of the Columbia Nanoscience Initiative. Additional processing was carried out at the PRISM Micro/Nano Fabrication Laboratory at Princeton University. The authors thank George P. Watson for technical assistance with the ICP-RIE etching performed there. The growth of GaAs/AlGaAs QWs at Purdue University was supported by grant DE-SC0006671 funded by the US Department of Energy Office of Science, Division of Materials Sciences and Engineering. The growth of GaAs/AlGaAs QWs at Princeton University was supported by the Gordon and Betty Moore Foundation under Award GMBF-2719 and by the National Science Foundation, Division of Materials Research, under Award DMR-0819860. VP acknowledges the European Graphene Flagship (contract No. CNECT-ICT-604391) for financial support and the Italian Ministry of Research (MIUR) through the program "Progetti Premiali 2012" - Project "ABNANOTECH".

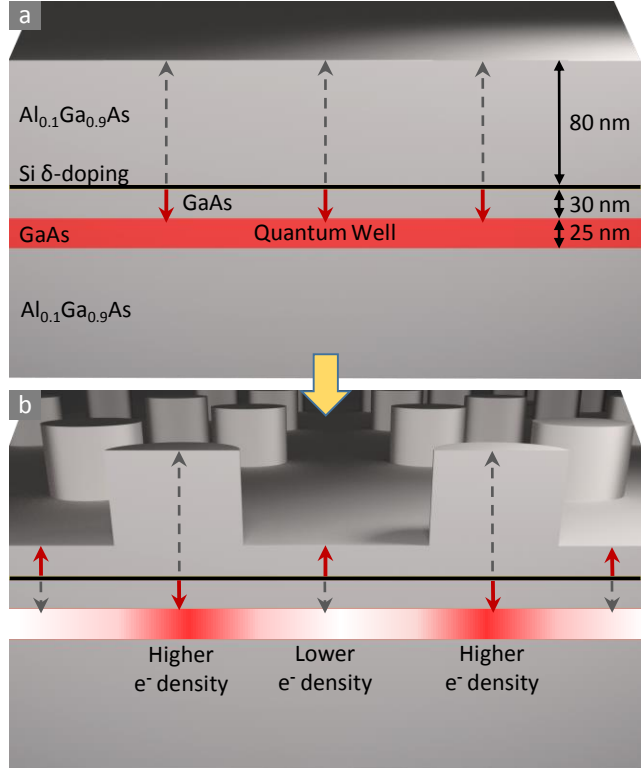

Figure 1. (Color online) Artificial graphene in a GaAs/AlGaAs heterostructure. (a) Schematic of the profile of the modulation-doped GaAs/Al$_{0.1}$Ga$_{0.9}$As quantum well. The 2DEG is confined by the quantum well potential within a 25 nm thick GaAs layer positioned 110 nm below the sample surface between two Al$_{0.1}$Ga$_{0.9}$As barriers. The Si impurities are inserted within the top barrier at a distance from the quantum well of 30 nm. The electrons from the Si donors predominantly occupy the ground state of the quantum well (red arrows), and to a lower extent the surface states (dashed gray arrows). (b) Artificial graphene created by etching into the Al$_{0.1}$Ga$_{0.9}$As barrier a honeycomb array of nano-pillars. The vicinity of the surface states to the doping layer results in a very efficient depletion of the 2DEG electron density in correspondence with the etched regions, symbolically represented by reversing the red and grey arrows and fading the red colored 2DEG, and quantum dots are formed beneath each pillar. The honeycomb lattice of interacting quantum dots produces energy minibands with linear dispersion, hosting MDFs.

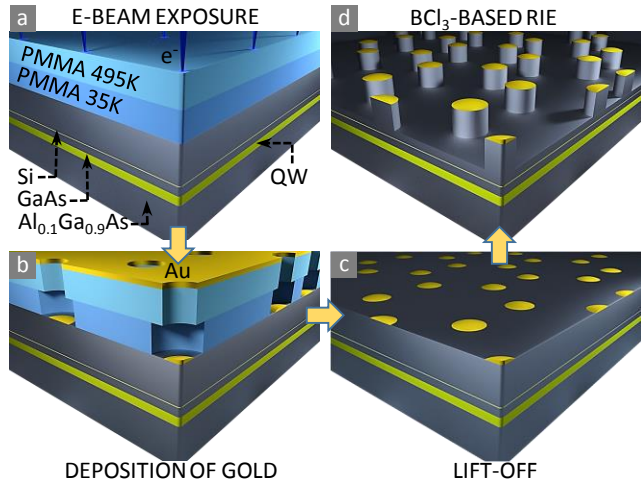

Figure 2. (Color online) Process flow for the creation of artificial graphene. (a) A PMMA bilayer resist stack is exposed by an electron beam at 80 kV accelerating voltage. The molecular weight of the polymer is indicated. (b) The development of the resist after exposure produces circular openings with an undercut profile. The metal constituting the etch mask (2 nm Ti + 8 nm Au) is deposited by e-beam evaporation perpendicular to the sample surface with high directionality. (c) The metal lift-off is accomplished by dissolving the resist in an NMP-based solution. A honeycomb array of gold nano-disks is hence produced on the sample surface. (d) The $BCl_3$-based ICP-RIE is performed to transfer the pattern to the substrate with high etching anisotropy and depth control.

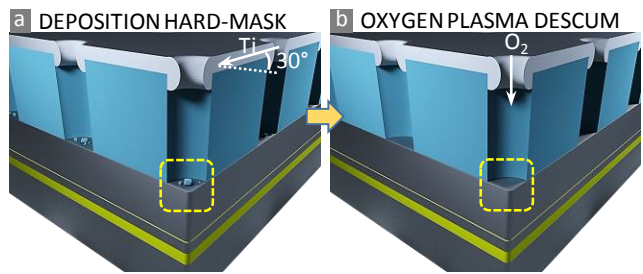

Figure 3. (Color online) Additional fabrication steps to facilitate the metal lift-off. (a) A metal hard-mask is formed on the resist by depositing a 5 nm thick layer of Ti at 60° angle with the vertical, which coats the PMMA surface without entering through the openings in the resist. The diameter of the developed features is reduced by the metal deposited along the top rim of the circular openings. (b) A directional RIE with oxygen plasma is applied in order to remove the residual resist and produce an undercut profile for easier lift-off. In addition, the $O_2$ plasma makes the surface hydrophilic increasing the adhesion of the metal mask.

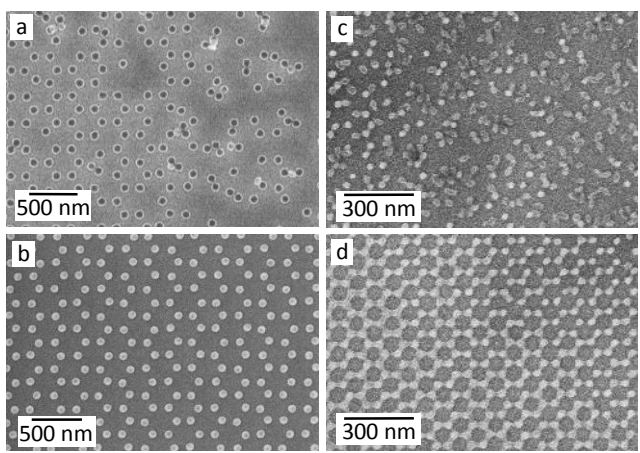

Figure 4. Honeycomb arrays of HSQ nano-posts. (a) Missing or misplaced sites due to poor adhesion to the substrate. The pitch is 150 nm and the posts diameter 75 nm. (b) Array with same parameters as in (a) on a substrate treated with the adhesion promoter Surpass 3000. (c) Honeycomb array with 50 nm lattice period and 25 nm nano-post diameter, on a substrate treated with adhesion promoter. Adhesion failure causes several defects. (d) Overexposed array with same parameters as in (c).

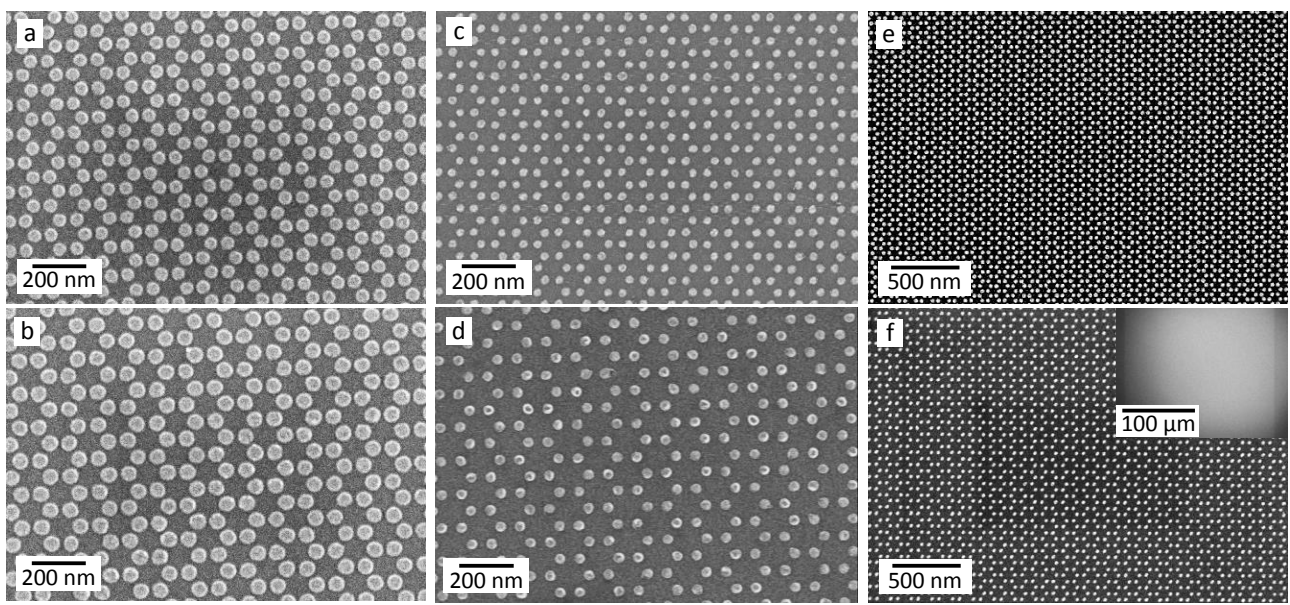

Figure 5. SEM micrographs of honeycomb arrays of Au nano-disks constituting the mask for the dry etching, with a variation in the lattice period and disk diameter. (a) 50 nm period and 40 nm diameter, (b) 70 nm period and 60 nm diameter, (c) 50 nm period and 25 nm diameter, (d) 70 nm period and 35 nm diameter. Lower magnification images in (e) and (f) show defect-free arrays with excellent uniformity over tens of unit cells for 40 nm and 50 nm period respectively. Inset: low magnification image of the entire 200 µm x 200 µm array.

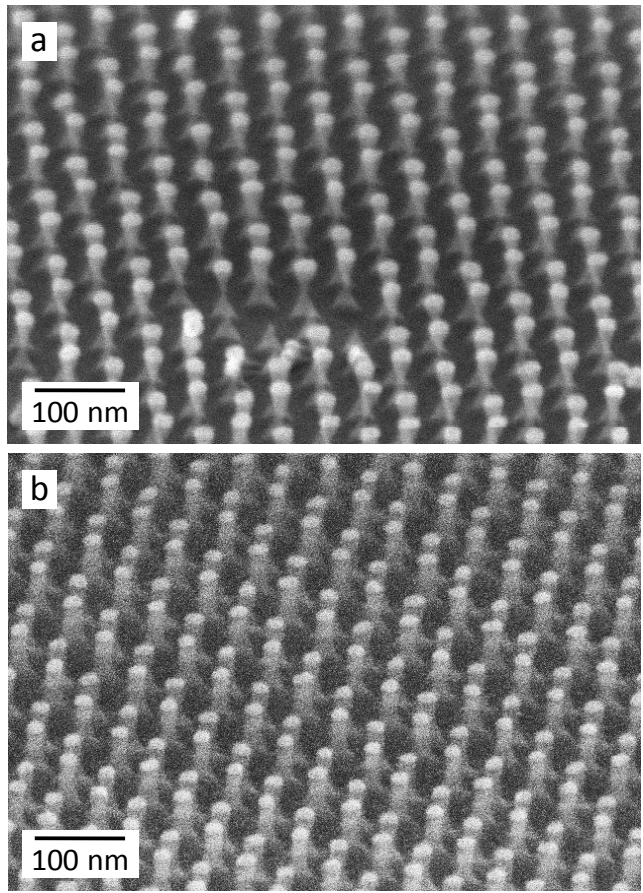

Figure 6. SEM micrographs (65° tilt) of AG lattices with 50 nm period for different etch conditions (disk diameter ∼ 20 nm; pillar height = 60 nm. Gas mixture: 50 sccm Ar, 5 sccm $BCl_3$; chamber pressure = 3.7 mbar). (a) RF power (RF2) = 50 W; ICP RF power (RF1) = 75 W. (b) RF2 = 75 W; RF1 = 50 W.

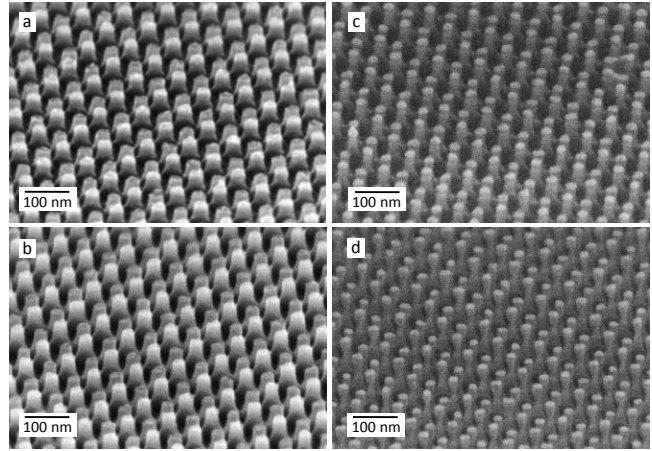

Figure 7. SEM micrographs (65° tilt) of artificial graphene lattices with 50 nm period for different etch times. (a) 60 s (etch depth = 45 nm); (b) 80 s (etch depth = 55nm); (c) 95 s (etch depth = 62 nm); (d) 110 s (etch depth = 75 nm). In (c) and (d) the metal mask is still present; this was taken into account for the evaluation of the depth of the etch. The sidewall profile is almost vertical in all cases, only the deepest etch presents a certain degree of undercut. The AG lattice in (d) was etched to within 5 nm from the Si doping layer, which constitutes the limit.

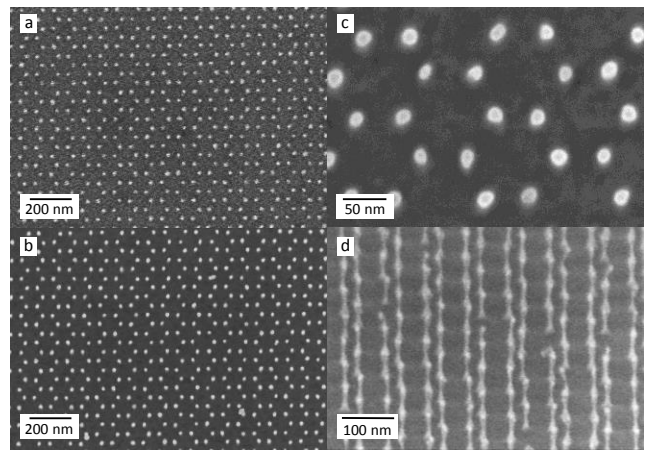

Figure 8. (a) SEM image of a 50 nm pitch honeycomb array of gold nano-disks obtained using the metal hard-mask method. (b) The same sample after annealing at 450 °C for two hours. The edge roughness of the Au disks noticeably improves. (c) Detail of the mask after annealing. The disk diameter is approximately 15 nm. (d) The resulting AG lattice after a 50 nm-deep etch with the optimized process (the metal mask is still present). The average width of the pillars is ~ 10 nm, resulting in an aspect ratio of 5.